# Exploring the Optimal Cycle for Quantum Heat Engine using Reinforcement Learning

Gao-xiang Deng[a], Haoqiang Ai[b], Bingcheng Wang[b], Wei Shao[a,b,*], Yu Liu[a], Zheng Cui[b,*]

a Institute of Thermal Science and Technology, Shandong University, Jinan, 250061, P.R. China

b Shandong Institute of Advanced Technology, Jinan, 250100, P.R. China

*Correspondence: shao@sdu.edu.cn; zhengc@sdu.edu.cn

**Abstract**

Quantum thermodynamic relationships in emerging nanodevices are significant but often complex to deal with. The application of machine learning in quantum thermodynamics has provided a new perspective. This study employs reinforcement learning to output the optimal cycle of quantum heat engine. Specifically, the soft actor-critic algorithm is adopted to optimize the cycle of three-level coherent quantum heat engine with the aim of maximal average power. The results show that the optimal average output power of the coherent three-level heat engine is 1.28 times greater than the original cycle (steady limit). Meanwhile, the efficiency of the optimal cycle is greater than the Curzon-Ahlborn efficiency as well as reporting by other researchers. Notably, this optimal cycle can be fitted as an Otto-like cycle, which illustrates the effectiveness of the method.

# 1 Introduction

The rapid advancement of science and technology has led to the miniaturization of devices, such as nano-process chips and nano-thermal engines [1]. Despite their small size, thermodynamic relationships within these microdevices, such as heat-dissipation and heat-work relationships (e.g., power, efficiency), remain crucial due to the quantum effects at the microscopic level. Quantum heat engines (QHEs) are devices that convert thermal energy to mechanical energy in a controlled way, using quantum-scale systems such as single particles or Qubits [2] as the working fluid. Studying QHEs can contribute to the emerging interdisciplinary field of quantum thermodynamics [3], elucidate the microscopic thermodynamic principles in miniaturized devices, and promote the development of nanotechnology [4].

In the investigation of QHEs, an open question that remains unresolved is whether quantum effects can be utilized to enhance their performance [5-10]. Recently, several QHEs have been constructed experimentally to investigate the aforementioned query. These approaches involve the manipulation of atomic spins [11, 12], ionic spins [13-15], or particle pair spins in crystals [16-18] through laser or magnetic field, the regulation of particles pairs spins utilizing nuclear magnetic resonance (NMR) technology [19-21], and the control of a single electron on a microcircuit cooled by dilution refrigeration [22-25]. These experiments achieved the cycle by means of state manipulation of the quantum system (working fluid) via electromagnetic pulse or voltage, with subsequent measurement of the state changes before and after the cycle to obtain the corresponding heat flux and power.

Although some positive conclusions, i.e. quantum effects enhanced the performance of QHEs, were reported [16-18, 20], the boosted performance typically requires a careful operation or specific condition. For example, the enhanced performance of QHE disappears when its thermal stroke time exceeds the decoherence time [16]. Moreover, when considering a specific cycle (e.g. Otto cycle), the impact of quantum effects on the performance becomes more ambiguous [10, 26-33]. This is primarily because these researches generally assume a specific thermodynamic cycle and this cycle may not ensure optimal power extraction on a long timescale.

Furthermore, prior theoretical researches on maximal power extraction usually focused on slow or fast driving regimes [34-40], assuming specific cycles [29, 41-47] such as the Otto cycle [29, 44-47], designing adiabatic shortcuts [48-54], or utilizing variational optimization [32, 55, 56]. The theoretical derivation and calculation of quantum thermodynamics can be extremely challenging, leading a need for numerous assumptions and a narrowed scope in these theoretical studies to obtain an analyzable solution. However, the utilization of reinforcement learning (RL) can potentially identify the optimal long-term power extraction cycles without such limitations, thereby may alleviating the need for tedious computations.

RL [57] have made significant progresses in various fields, including computer games [58-60], robotics [61], and natural language processing [62]. These algorithms exhibit a much stronger exploration ability than humans and have been used for the quantum state preparation and quantum computing [63-71], surpassing the traditional methods used before. What's more, RL algorithms have been applied to explore the optimal cycle for two-level QHE and harmonic oscillator QHE [72-74]. Despite the potential of RL algorithms in many-body or multi-level quantum systems, their application has been relatively unreported due to the lack of proper dynamical evolution modeling that simplifies theoretical analysis.

This study employs the RL by the soft actor-critic (SAC) algorithm aiming to explore the optimal cycles with maximal long-term performance for coherent three-level QHE. Subsequently, the convergence of the SAC algorithm is analyzed through five consecutive trainings and the power and efficiency are discussed. Finally, applying the Boltzmann function during the compression and expansion processes approximates the optimal cycle as an Otto-like cycle.

## 2 Models and method

### 2.1 Thermodynamic model of coherent three-level QHE

The thermal processes of the coherent three-level QHE is governed by the dynamics of the transitions between the energy levels of the quantum system. Fig. 1 depicts the thermodynamic model of the coherent three-level QHE. Fig. 1(a) shows three energy levels, i.e., $|0\rangle$, $|1\rangle$, and $|2\rangle$, of the quantum system. The transition from $|0\rangle$ to $|2\rangle$, from $|1\rangle$ to $|0\rangle$, or between $|1\rangle$ and $|2\rangle$ occurred when the quantum system couples with a hot reservoir at temperature $T_h$, a cold reservoir at temperature $T_c$, and an external field $V$, respectively. Fig. 1(b) illustrates that the coupling to the hot reservoir, the cold reservoir, and the external field will lead to heat absorption $Q_h$, heat release $Q_c$, and work output $W$, respectively.

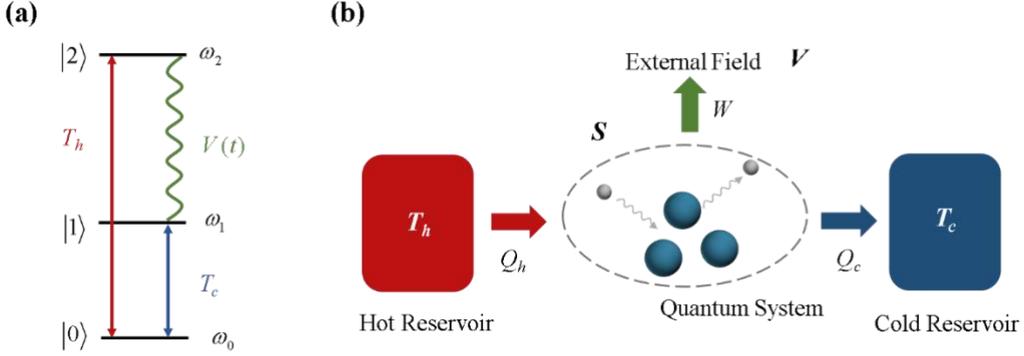

Fig. 1. Schematic of coherent three-level QHE. (a) Energy levels. $|0\rangle$, $|1\rangle$, and $|2\rangle$ are the eigenstates of the quantum system's free Hamiltonian, and $\omega_0$, $\omega_1$, and $\omega_2$ are the corresponding eigenfrequencies. The thermodynamics between these energy levels are given in the main text. (b) Thermal processes. The quantum system undergoes three different processes: absorbing heat $Q_h$ from the hot reservoir at temperature $T_h$, releasing heat $Q_c$ to the cold reservoir at $T_c$, and the outputting work $W$ to the external field $V(t) = \lambda e^{i\omega t}|1\rangle\langle 2| + \lambda e^{-i\omega t}|2\rangle\langle 1|$. $\lambda$, $\omega$, and $t$ are the intensity, the frequency, and the evolution time of the external field, respectively.

The quantum system $S$ in this coherent three level QHE is governed by the Gorini-Kossakowski-Lindblad-Sudarshan (GKLS) equation [75, 76],

$$\partial_t \rho_S = -\frac{i}{\hbar}[H_S, \rho_S] + \sum_{i=c,h} D_i[\rho_S(t)] \qquad (1)$$

where $H_S$ denotes the Hamiltonian of the quantum system $S$, $D_i$ is the dissipator which represents the heat-dissipation to the reservoirs, the subscripts, $c$ and $h$, represent the cold and hot reservoir, respectively. The corresponding dissipator is defined as:

$$D_i = \sum_{\varepsilon} \Gamma_i(\varepsilon) \left( L_i^{\varepsilon}(t) \rho_S \left[ L_i^{\varepsilon}(t) \right]^{\dagger} - \frac{1}{2} \left\{ \left[ L_i^{\varepsilon}(t) \right]^{\dagger} L_i^{\varepsilon}(t) \rho_S + \rho_S \left[ L_i^{\varepsilon}(t) \right]^{\dagger} L_i^{\varepsilon}(t) \right\} \right) \quad (2)$$

and the projected jump operator is:

$$L_i^{\varepsilon}(t) = \left[ L_i^{-\varepsilon}(t) \right]^{\dagger} = \sum_{m,n=0}^{2} \delta_{\varepsilon, \varepsilon_m - \varepsilon_n} \left| \varepsilon_n(t) \right\rangle \left\langle \varepsilon_n(t) \right| L_i \left| \varepsilon_m(t) \right\rangle \left\langle \varepsilon_m(t) \right| \quad (3)$$

with $L_c = |0\rangle\langle 1|$, $L_h = |0\rangle\langle 2|$. Here, the coupling parameter satisfies:

$$\Gamma_i(-\varepsilon) = e^{-\beta_i \varepsilon} \Gamma_i(\varepsilon) \quad (4)$$

where, $\beta_i = 1/k_B T_i$ is the inversed temperature of the reservoir, $\varepsilon$ and $|\varepsilon(t)\rangle$ are the eigenvalues and eigenstates of $H_S$, respectively. In the following sections, both the Boltzmann constant $k_B$ and reduced Planck constant $\hbar$ are set as 1.

### 2.2 RL model of coherent three-level QHE

This section presented the RL model of the coherent three-level QHE based on the thermodynamic model, which begins with outlining the basic setting for the RL model, followed by the description of the long-term performance, the reward function, and the training details.

### 2.2.1 Basic settings for the RL model

Fig. 2(a) demonstrates a coherent three-level QHE whose evolution is controlled by an RL agent. The objective of the RL agent is to identify the cycle that maximizes the long-term performance of the QHE by optimizing both the discrete control parameter, $d(t) = \{\text{hot, cold, work}\}$, and the continuous control parameter, $u(t)$. It is noteworthy that the optimized combinations of $a(t) = \{d(t), u(t)\}$ at different times are the cycles that maximizes the long-term performance.

The discrete control parameter $d(t)$ determines the thermal process, with the settings for different thermal processes provided in Table 1 and Table 2. The frequency difference between energy level $|1\rangle$ and $|0\rangle$ is chosen as reference and is written as follows:

$$\omega_{10} = \omega_1 - \omega_0 \quad (5)$$

Similarly, $\omega_{20} = \omega_2 - \omega_0$.

Table 1. Parameters of the reservoirs and external field for the coherent three-level QHE in different processes. $g_1$ and $g_2$ are coupling functions, which respectively represent the coupling to the cold and hot reservoir, $\varepsilon_{21} = \varepsilon_2 - \varepsilon_1$.

| Process | $\beta_c \omega_{10}$ | $\beta_h \omega_{10}$ | $\omega_{20}/\omega_{10}$ | $\lambda/\omega_{10}$ | $\omega$ [76] |
|---|---|---|---|---|---|
| Hot | 0 | 1 | 2.5 | 0 | 0 |
| Cold | 5 | 0 | 2.5 | 0 | 0 |
| Work | 0 | 0 | 2.5 | 0.5 | $\dfrac{\varepsilon_{21}^2 + \dfrac{1}{4}(g_1+g_2)^2}{\omega_2 - \omega_1}$ |

Table 2. Coupling parameters of the coherent three-level QHE in different processes.

| Process | $\Gamma_c(\varepsilon_{10})/\omega_{10}$ | $\Gamma_h(\varepsilon_{20})/\omega_{10}$ | $\Gamma_c(\varepsilon_{20})/\omega_{10}$ | $\Gamma_h(\varepsilon_{10})/\omega_{10}$ |
|---|---|---|---|---|
| Hot | 0 | 2 | 0 | 0 |
| Cold | 2 | 0 | 0 | 0 |
| Work | 0 | 0 | 0 | 0 |

The continuous control parameter, denoted as $u(t)$, is initialized at the onset of each thermal process ($t$) and remains constant until the process concludes ($t+\Delta t$). This parameter dictates the free Hamiltonian and energy gaps of the quantum system, thereby indicating how the agent manipulates the states of the quantum system throughout each thermal process. The Hamiltonians of the quantum system $S$ at $t$ and $t+\Delta t$ can be denoted, respectively, as

$$H_S(t) = u(t) H_{\text{free}} + V(0) = \begin{pmatrix} u(t)\omega_0 & 0 & 0 \\ 0 & u(t)\omega_1 & \lambda \\ 0 & \lambda & u(t)\omega_2 \end{pmatrix} \qquad (6)$$

$$H_S(t+\Delta t) = u(t) H_{\text{free}} + V(\Delta t) = \begin{pmatrix} u(t)\omega_0 & 0 & 0 \\ 0 & u(t)\omega_1 & \lambda \cdot e^{i\omega \Delta t} \\ 0 & \lambda \cdot e^{-i\omega \Delta t} & u(t)\omega_2 \end{pmatrix} \qquad (7)$$

where the non-zero value of the non-diagonal elements in the Hamiltonians induces coherence within the QHE. Here,

$$H_{\text{free}} = \begin{pmatrix} \omega_0 & 0 & 0 \\ 0 & \omega_1 & 0 \\ 0 & 0 & \omega_2 \end{pmatrix} \tag{8}$$

is the free Hamiltonian, and

$$V(\tau) = \lambda e^{i\omega\tau} |1\rangle\langle 2| + \lambda e^{-i\omega\tau} |2\rangle\langle 1| = \begin{pmatrix} 0 & 0 & 0 \\ 0 & 0 & \lambda e^{i\omega\tau} \\ 0 & \lambda e^{-i\omega\tau} & 0 \end{pmatrix}, \tau \in [0, \Delta t] \tag{9}$$

$$V(\tau + \Delta t) = V(\tau)$$

This implies that the external field is initialized at the commencement of each thermal process to avert phase accumulation across distinct processes. As per the configurations in Table 1, the external field is activated solely during the work process, while it is deactivated during both the hot and cold processes.

As long as $d(t)$ and $u(t)$ are given by the RL agent, the evolution of $S$ from $t$ to $t + \Delta t$ can be obtained by solving Eq. (1). By applying Eq. (6) and Eq. (7), and substituting the corresponding parameter values for different thermal processes as provided in Table 1 and Table 2, the value of $\rho_S(t + \Delta t)$ can be determined.

Fig. 2(b) illustrates the RL training processes of the coherent three-level QHE shaded in green, while the RL agent, modeled by the NNs updating through the SAC algorithm shaded in blue. The quantum system takes an action $a_t$ based on the policy provided by the RL agent then transitions to a new state $s_{t+1}$ and receives a reward $r_{t+1}$. Here, the thermodynamic model of the coherent QHE is designed to generate state transitions and rewards for the actions provided by the RL agent.

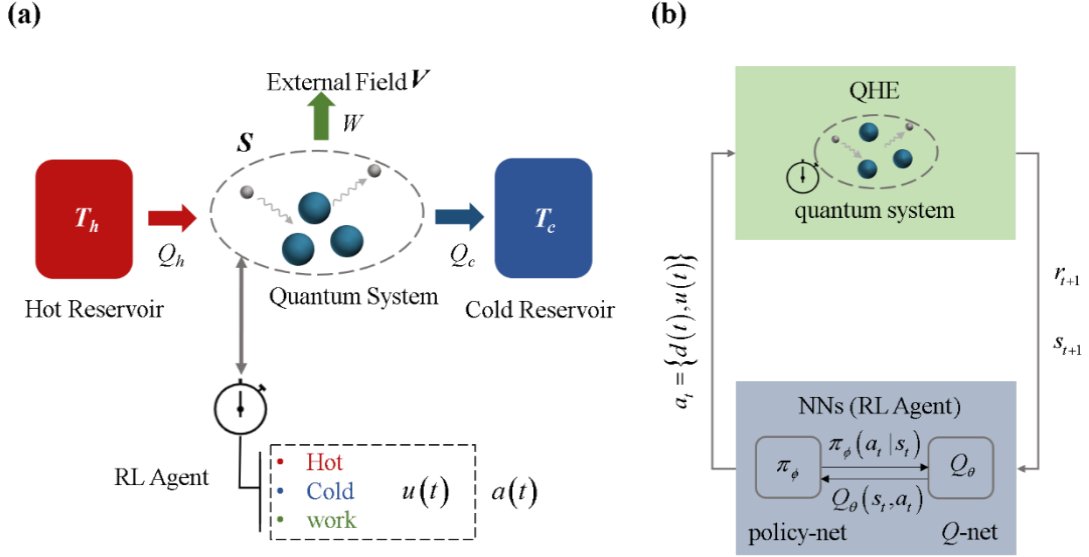

Fig. 2. Schematic of applying RL algorithm to optimize the long-term performance of a coherent three-level QHE. (a) The RL agent controls the evolution of the QHE through $d(t) = \{\text{hot, cold, work}\}$, which determines the thermal process, and $u(t)$, which determines the corresponding system state. $a(t) = \{d(t), u(t)\}$ is a composite control parameter of $d(t)$ and $u(t)$. The specific definitions of $d(t)$ and $u(t)$ are provided in the main text. (b) RL training processes of the coherent three-level QHE. "QHE" refers to the thermodynamic model of QHE, representing the evolution of the quantum system with the control of the RL agent. "NNs" are the neural networks of the RL agent, which are composed of Q-net $Q_\theta$ and policy-net $\pi_\phi$. The quantum system takes an action $a_t$ based on the policy given by the RL agent then transitions to a new state $s_{t+1}$ while receiving a reward $r_{t+1}$. The NNs receive $s_{t+1}$ and $r_{t+1}$ as input then output a new action based on a new policy. These steps are repeated until convergence to the optimal policy, which produces the optimal cycle with the maximal long-term performance. See Section 2.3 for more details.

### 2.2.2 Long-term performance

The output power $P_{he}$ are generally used as the performance metric for heat engines,

$$P_{he} = \sum_{\alpha=c,h} J_\alpha(t) \tag{10}$$

where, $J$ is the heat current; and, the subscript $c$ and $h$ refer to the cold reservoir and hot reservoir, respectively. However, the aim of this research is to obtain the optimal cycle with maximal long-term performance. Hence, the average power [72],

$$\langle P_{he} \rangle = \tilde{\gamma} \int_0^\infty e^{-\tilde{\gamma}t} P_{he}(t)\, dt \tag{11}$$

was chosen as the long-term performance metric. The timescale of interest can be manipulated by adjusting $\tilde{\gamma}$, with smaller $\tilde{\gamma}$ and larger $\tilde{\gamma}$ correspond to longer and shorter timescale, respectively.

### 2.2.3 Reward function

In reinforcement learning, rewards are generally returned through a designed reward function. The designed reward function needs to possess a clear physical meaning and ensure convergence as well as maximizing the average power in Eq. (11).

Therefore, according to Eq. (10) and (11), the reward function for the RL model of coherent three-level QHE should be,

$$r_{QHE} = \delta_{d,\bar{d}} \frac{1}{\Delta t} \Delta \langle E_S \rangle \tag{12}$$

which represents the average internal energy change rate of quantum system $S$ change from $t$ to $t+\Delta t$. Here, $d = \{\text{hot, cold, work}\}$, $\bar{d} = \{\text{hot, cold}\}$, $\Delta t$ is the time step,

$$\delta_{d,\bar{d}} = \begin{cases} 1, & d = \text{hot or cold} \\ 0, & d = \text{work} \end{cases} \tag{13}$$

and,

$$\Delta \langle E_S \rangle = \langle E_S(t+\Delta t) \rangle - \langle E_S(t) \rangle \tag{14}$$

is the change of internal energy $\langle E \rangle = \text{tr}(\rho H)$.

The optimization of average power can be regarded as a discounted RL task [57, 77], which operates in both continuous and discrete spaces. This approach has been demonstrated to be effective in learning far-from-equilibrium finite-time thermodynamics [69, 72]. Specifically, the RL agent aims to maximize the total future reward [57, 78],

$$r_{i+1} + \gamma r_{i+2} + \gamma^2 r_{i+3} + \cdots = \sum_{k=0}^\infty \gamma^k r_{i+1+k} \tag{15}$$

where, $i+1$ is the next step, $\gamma \in [0,1)$ is the discount factor with smaller $\gamma$ and larger $\gamma$ correspond to shorter and longer future reward, respectively. According to Eq. (14), the reward $r_{i+1}$ is given by,

$$r_{i+1} = \delta_{d,\bar{d}} \frac{1}{\Delta t} \Delta \langle E_S \rangle = \delta_{d,\bar{d}} \frac{1}{\Delta t} \left[ \text{tr}\left(\rho_S(t+\Delta t) H_S(t+\Delta t)\right) - \text{tr}\left(\rho_S(t) H_S(t)\right) \right] \quad (16)$$

where,

$$t = i\Delta t \quad (17)$$

Substitute Eq. (16) into Eq. (15) comes to the conclusion that the aim of the RL agent is to maximize Eq. (11) with $\tilde{\gamma} = -\ln\gamma/\Delta t$ [72]. Consequently, the future average power, $\langle P_{he} \rangle$, in Eq. (11) and the average power of the current step $i$, $\langle P_{he} \rangle_i$, can be, respectively, expressed as:

$$\langle P_{he} \rangle = (1-\gamma) \sum_{k=0}^{\infty} \gamma^k r_{i+1+k} \quad (18)$$

$$\langle P_{he} \rangle_i = (1-\gamma) \sum_{k=0}^{i} \gamma^k r_{i-k} \quad (19)$$

It can be demonstrated that Eq. (18) and (19) exhibit consistent convergence, indicating that they converge to the same value. This implies that maximizing either Eq. (18) or (19) will yield the same effect. Thus, it turns out that the aim of the RL agent is to find the optimal cycle that maximizes the average power defined in Eq. (11).

**2.2.4 Training details**

The training parameters of RL for the coherent three-level QHE are listed in Table 3.

Table 3. Training parameters

| Parameter | Coherent three-level QHE |
|---|---|
| optimizer | Adam |
| learning rate | $3\times10^{-4}$ |
| number of hidden layers | 2 |
| number of hidden units per layer | 256 |
| activation function | ReLU |
| size of buffer $R$ | $160\times10^3$ |
| batch size | 512 |
| discount $\gamma$ | 0.995 |
| time step $\Delta t$ | 0.5 |
| "polyak" coefficient $\tau$ | 0.005 |
| update steps | 50 |
| lower and upper limit of $u$ | [0.3, 1.5] |
| $\bar{H}_{D,\text{init}}$ | $0.98\times\ln3$ |
| $\bar{H}_{D,\text{final}}$ | 0.03 |
| $\bar{H}_{D,\text{decay}}$ | $144\times10^3$ |
| $\bar{H}_{C,\text{init}}$ | $-0.72$ |
| $\bar{H}_{C,\text{final}}$ | $-3.8$ |
| $\bar{H}_{C,\text{decay}}$ | $144\times10^3$ |
| training steps | $500\times10^3$ |

**2.3 SAC algorithm of RL**

The SAC algorithm is a type of Actor-Critic (AC) RL algorithm that can be divided into three parts: optimization objective, policy evaluation, and policy improvement. The SAC algorithm improves its stability and exploration by introducing an entropy term in the optimization objective [78, 79]. As for the iteration steps, firstly, policy evaluation, the actor in this algorithm selects an action based on a given policy, and the critic scores this selected action. Subsequently, policy improvement, the actor improves its policy based on the scores from the critic. The "policy" which takes the form of a probability refers to how the actor chooses actions and "scores" from the critic can be expressed as $Q$ function.

In order to avoid the ambiguity, it is necessary to note that the following symbols and terms have no relation to those used in the preceding text. For instance, the symbol "$H$" in the SAC algorithm refers to the entropy of a specific policy, not the Hamiltonian.

### 2.3.1 Optimization objective

The optimization objective can be expressed as:

$$\pi^* = \arg\max_{\pi} \sum_{t=0}^{\infty} \mathrm{E}_{(s_t, a_t) \sim \rho_{\pi}} \left[ r(s_t, a_t) + \alpha \cdot H(\pi(\cdot|s_t)) \right] \tag{20}$$

where, $\rho_{\pi}$ is the state-action marginals of the trajectory distribution induced by a policy $\pi$. The current reward obtained by the actor taking an action $a_t$ in the state $s_t$ is denoted as $r(s_t, a_t)$ or $r_t$ for simplicity. Additionally, $H$ is the entropy of a probability distribution, and $\pi(\cdot|s_t)$ represents the probability distribution of choosing an arbitrary action "$\cdot$" in the state $s_t$. The temperature $\alpha$ represents the weight of entropy in the expected reward.

### 2.3.2 Policy evaluation

The SAC algorithm employs $Q$ to evaluate the policy $\pi$:

$$Q^{\pi} = \mathrm{E}_{(s_t, a_t) \sim \rho_{\pi}} \left\{ \sum_{t=0}^{\infty} \gamma^t \cdot r(s_t, a_t) + \sum_{t=1}^{\infty} \gamma^t \cdot \alpha H(\pi(\cdot|s_t)) \Big| s_0 = s, a_0 = a \right\} \tag{21}$$

Here, $\gamma \in [0,1)$ is the discount factor with the consideration that the weight of the early training data should be reduced, and $s_0$ ($a_0$) is the initial state (action). The value of $Q$ will be converged through iteration according to the Bellman backup operator,

$$\mathrm{B}^{\pi} Q(s_t, a_t) \triangleq r(s_t, a_t) + \gamma \cdot \mathrm{E}_{s_{t+1} \sim R} \left\{ \mathrm{E}_{a' \sim \pi} \left[ Q(s_{t+1}, a') - \alpha \log \pi(a'|s_{t+1}) \right] \right\} \tag{22}$$

where, $R$ is the replay buffer which is used to replace the unknown distribution $\rho_{\pi}$, $s_{t+1}$ is the state of next step, and $a'$ is the action of next step (improved action). The iteration based on Bellman backup between training step $k$ and step $k+1$ is:

$$Q_{k+1} = \mathrm{B}^{\pi} Q_k \tag{23}$$

The SAC algorithm adopts two NNs with parameters $\theta_1$ and $\theta_2$ to fit the $Q$ function, which is the "double Q-learning trick" for continuous control and other improvements [78]. These NNs are generally called $Q$-net and are denoted as $Q_{\theta_j}$, with $j = \{1, 2\}$. Specifically, the inputs of $Q_{\theta_j}$ are ($s$, $a$) and the outputs are the value of $Q$ function. The outputs of $Q_{\theta_j}$ will be converged by minimizing the loss function,

$$L_Q(\theta_j) = \mathrm{E}_{\substack{(s_t, a_t, r_t, s_{t+1}) \sim R \\ a' \sim \pi}} \frac{1}{2}\left[Q_{\theta_j}(s_t, a_t) - y(r_t, s_{t+1})\right]^2 \qquad (24)$$

where,

$$\begin{aligned} y(r_t, s_{t+1}) &= \mathrm{B}^\pi Q_{\bar{\theta}}(s_t, a_t) \\ &= r(s_t, a_t) + \gamma \cdot \mathrm{E}_{s_{t+1} \sim R}\left\{\mathrm{E}_{a' \sim \pi}\left[\min_{j=1,2} Q_{\bar{\theta}_j}(s_{t+1}, a') - \alpha \log \pi(a'|s_{t+1})\right]\right\} \end{aligned} \qquad (25)$$

and, $\bar{\theta}_j$ is the parameters of the target $Q$-net $Q_{\bar{\theta}_j}$. Here, the parameters of $Q_{\bar{\theta}_j}$ are not updated during the backpropagation but are updated through "polyak", i.e.

$$\bar{\theta}_j \leftarrow \tau \theta_j + (1-\tau)\bar{\theta}_j \qquad (26)$$

to improve learning [78]. Here, $\tau$ is a hyperparameter and is listed in Table 3.

The iteration step of policy evaluation is shaded in green in Fig. 3. The Q-net $Q_{\theta_j}$ and the target Q-net $Q_{\bar{\theta}_j}$ take data fetched from the replay buffer as input to output corresponding values then update $\theta_j$ and $\bar{\theta}_j$ by minimizing the loss function $L_Q(\theta_j)$ given in Eq. (24) and by (26), respectively.

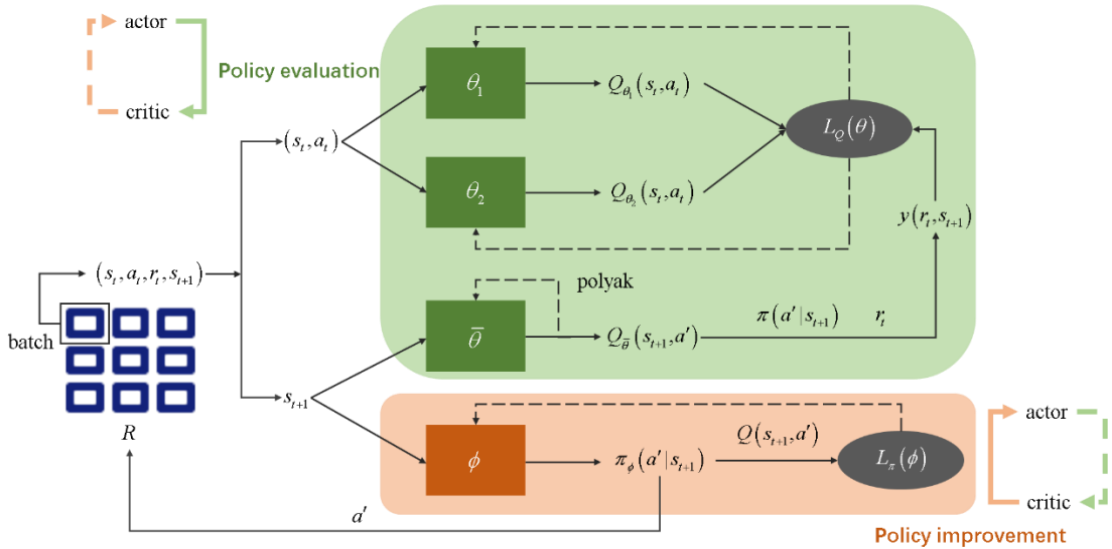

Fig. 3. Steps of SAC algorithm. A batch-sized data, which includes the current state

$s_t$, current action $a_t$, current reward $r_t$, and next state $s_{t+1}$, was firstly fetched from the replay buffer $R$ for the training. The training of SAC algorithm can be divided into two main steps: policy evaluation and policy improvement. (1) Policy evaluation (green). First, $(s_t, a_t)$ are inputted into Q-net whose parameters are $\theta_1$ and $\theta_2$, and outputting $Q_{\theta_1}(s_t, a_t)$ and $Q_{\theta_2}(s_t, a_t)$ to evaluate the current action $a_t$. After that, $Q_{\bar{\theta}}(s_{t+1}, a')$ is obtained by inputting $s_{t+1}$ and the next action $a'$ given by the policy-net into the target Q-net whose parameters are $\bar{\theta}$. Subsequently, substitute policy $\pi(a'|s_{t+1})$, $r_t$, and $Q_{\bar{\theta}}(s_{t+1}, a')$ into Eq. (25) to obtain $y(r_t, s_{t+1})$, and yield the loss function of Q-net $L_Q(\theta)$ through Eq. (24). Lastly, the gradient of $L_Q(\theta)$ is backpropagated (dotted arrow) to update $\theta_1$ and $\theta_2$, and $\bar{\theta}$ is "soft" updated through "polyak" (dotted arrow). (2) Policy improvement (orange). First, the policy-net receive the next state $s_{t+1}$ and output the policy $\pi_\phi(a'|s_{t+1})$, then the next (improved) action $a'$ is sampled from $\pi_\phi(a'|s_{t+1})$. Subsequently, substitute $\pi_\phi(a'|s_{t+1})$ and $Q(s_{t+1}, a')$ outputted by the Q-net into Eq. (29) gives the loss function of policy-net $L_\pi(\phi)$, which was used lastly in backpropagation (dotted arrow) to update $\phi$.

### 2.3.3 Policy improvement

The policy improvement step in the RL algorithm is implemented to obtain the optimal policy $\pi^*$. Ideally, the policy obeys the following probability distribution form:

$$\begin{aligned} \pi(a_t | s_t) &\sim \exp(-\varepsilon(s_t, a_t)) \\ \varepsilon(s_t, a_t) &= -\frac{1}{\alpha} Q(s_t, a_t) \end{aligned} \tag{27}$$

Eq. (27) is the energy-based distribution function for complex tasks, but it cannot be sampled or provide a specific action. Therefore, the Gaussian distribution, denoted by $\pi$, is used to approximate the above energy-based distribution. This approximated probability distribution $\pi$ should minimize the Kullback-Leibler (KL) divergence,

$$\pi_{new} = \arg\min_{\pi \sim \Pi} D_{KL}\left[ \pi(\cdot|s_t) \middle\| \frac{\exp(1/\alpha \cdot Q^{\pi_{old}}(s_t, a_t))}{Z^{\pi_{old}}(s_t, a_t)} \right] \tag{28}$$

where, $D_{KL}(P\|Q) = -\sum_i p_i \ln(q_i/p_i)$, $\Pi$ represents all possible sets of policy $\pi$,

and *Z* is a normalization constant which can be ignored during the backpropagation. Consequently, the loss function of policy neural network (policy-net for simplicity) $\pi_\phi$ can be obtained as:

$$L_\pi(\phi) = \mathop{\mathrm{E}}_{\substack{s_{t+1} \sim R \\ a' \sim \pi_\phi}} \left[ \alpha \cdot \log \pi_\phi(a' | s_{t+1}) - Q(s_{t+1}, a') \right] \tag{29}$$

where, $\phi$ is the set of the parameters of $\pi_\phi$. Sampling from the Gaussian distribution $\pi(a' | s_{t+1})$ will produce the improved action $a'$ for the next state $s_{t+1}$. The orange area in Fig. 3 illustrates the processes of policy improvement step.

**2.3.4 Modifications of SAC algorithm**

Given that the SAC algorithm is designed for continuous action space. The optimization in this research is carried out both in discrete and continuous spaces. Therefore, it becomes necessary to make the following modifications to the SAC algorithm.

The temperature $\alpha$ is separated into two components: $\alpha_D$ for the discrete action $d(t)$ and $\alpha_C$ for the continuous action $u(t)$. The values of $\alpha_D$ and $\alpha_C$ can be updated by minimizing the following loss functions [73, 78]:

$$\begin{aligned} L_{\alpha_D}(\alpha_D) &= \alpha_D \cdot \mathrm{E}_{s \sim R}\left[ H_D^\pi(s) - \bar{H}_D \right] \\ L_{\alpha_C}(\alpha_C) &= \alpha_C \cdot \mathrm{E}_{s \sim R}\left[ H_C^\pi(s) - \bar{H}_C \right] \end{aligned} \tag{30}$$

where,

$$\begin{aligned} H_D^\pi(s) &= -\sum_d \pi_D(d|s) \log \pi_D(d|s) \\ H_C^\pi(s) &= -\sum_d \pi_D(d|s) \mathrm{E}_{u \sim \pi_C(\cdot | d, s)}\left[ \log \pi_C(u | d, s) \right] \end{aligned} \tag{31}$$

is the current entropy of the discrete policy $\pi_D$ and the continuous policy $\pi_C$, respectively. And,

$$\begin{aligned} \bar{H}_D &= \bar{H}_{D,\mathrm{final}} + \left( \bar{H}_{D,\mathrm{init}} - \bar{H}_{D,\mathrm{final}} \right) \exp\left( -n_{\mathrm{steps}} / \bar{H}_{D,\mathrm{decay}} \right) \\ \bar{H}_C &= \bar{H}_{C,\mathrm{final}} + \left( \bar{H}_{C,\mathrm{init}} - \bar{H}_{C,\mathrm{final}} \right) \exp\left( -n_{\mathrm{steps}} / \bar{H}_{C,\mathrm{decay}} \right) \end{aligned} \tag{32}$$

is the target entropy of discrete and continuous policy, respectively, $n_{\mathrm{steps}}$ is the current trained step. It is noteworthy that both $\alpha_D$ and $\alpha_C$ are the parameter and the output of the NNs. Consequently, they are typically initialized as zero-dimensional tensors before training.

Replacing $E_{a'\sim\pi}[-\alpha\log\pi(a'|s_{t+1})]$ in Eq. (25) with $\alpha_D H_D^\pi(s_{t+1})+\alpha_C H_C^\pi(s_{t+1})$ yields the loss function for the Q-net:

$$L_Q(\theta_j) = E_{\substack{(s_t,a_t,r_t,s_{t+1})\sim R \\ a'\sim\pi}} \frac{1}{2}\left[Q_{\theta_j}(s_t,a_t) - y(r,s_{t+1})\right]^2 \quad (33)$$

Here,

$$y(r,s_{t+1}) = r(s_t,a_t) \\ + \gamma \cdot E_{s_{t+1}\sim R}\left\{E_{a'\sim\pi}\left[\min_{j=1,2} Q_{\bar{\theta}_j}(s_{t+1},a')\right] + \alpha_D H_D^\pi(s_{t+1}) + \alpha_C H_C^\pi(s_{t+1})\right\} \quad (34)$$

and, the value of target Q-net is [77, 80]:

$$E_{a'\sim\pi}\left[\min_{j=1,2} Q_{\bar{\theta}_j}(s_{t+1},a')\right] = \sum_{d'} \pi_D(d'|s_{t+1}) E_{u'\sim\pi_C(\cdot|d',s_{t+1})}\left[\min_{j=1,2} Q_{\bar{\theta}_j}(s_{t+1},d',u')\right] \quad (35)$$

The entropy of the current policy is:

$$\alpha_D H_D^\pi(s_{t+1}) + \alpha_C H_C^\pi(s_{t+1}) = -\alpha_D \sum_{d'} \pi_D(d'|s_{t+1})\log\pi_D(d'|s_{t+1}) \\ -\alpha_C \sum_{d'} \pi_D(d'|s_{t+1}) E_{u'\sim\pi_C(\cdot|d',s_{t+1})}\left[\log\pi_C(u'|d',s_{t+1})\right] \quad (36)$$

Similarly, replacing $E_{a_t\sim\pi_\phi}[\alpha\cdot\log\pi_\phi(a_t|s_t)]$ in Eq. (29) with $\alpha_D H_D^{\pi_\phi}(s) + \alpha_C H_C^{\pi_\phi}(s)$ gives the loss function of the policy-net:

$$L_\pi(\phi) = E_{\substack{s_t\sim R \\ a_t\sim\pi_\phi}}\left[\alpha\cdot\log\pi_\phi(a_t|s_t) - Q_\theta(s_t,a_t)\right] \\ = E_{s_t\sim R}\left[\sum_d \pi_{D,\phi}(d|s_t)\alpha_D\log\pi_{D,\phi}(d|s_t) + \alpha_C\log\pi_{C,\phi}(u|d,s_t) \\ - \sum_d \pi_{D,\phi}(d|s_t)\min_{j=1,2} Q_{\theta_j}(s_t,d,u)\right] \quad (37)$$

## 3 Results and Discussions

### 3.1 Training results

The average output power $\langle P_{he} \rangle$ and the policies during the training given in Fig. 4(a) and 4(b), respectively. Fig. 4(a) demonstrates that the maximum average output power of "RL Cycle" (solid line) converges to about 0.91, which is approximately 1.28 times greater than 0.399 of the "Steady" limit (dashed line). The reason is that the agent designs different $u$ for different processes $d$, strengthening or weakening the corresponding processes to improve the average output power. Additionally, Fig. 4(b) shows the convergence of different policies (cycles) provided by the RL agent as the number training steps increases.

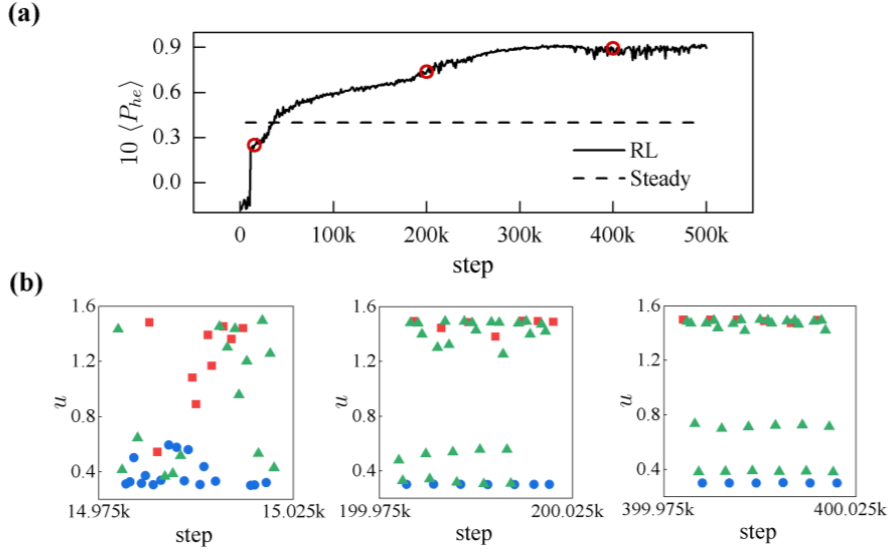

Fig. 4. Training results of the entangled three-level QHE. (a) Average output power $\langle P_{he} \rangle$ during the training. The solid line represents the average output power obtained by the RL agent and the dashed line represents the "Steady" limit derived in ref. [76]. One step corresponds to time step $\Delta t$ in Table 3. (b) Different policies (cycle) given by the RL agent during different training periods marked in figure (a). The red, blue, and green points represent hot, cold, and work process, respectively, and $u$ denotes the corresponding system state of these processes. See Table 3 for detail training parameters.

The well-trained RL agent produces the optimal cycle as shown in Fig. 5. The optimal $u$ for the hot and cold process are fixed at 1.5 and 0.3, respectively. However, optimal $u$ for the work processes varies with time between 0.4 and 1.5. At the same time, the corresponding process time steps for the hot, cold, and work process are 1, 1, and 5, respectively. A further discussion on these results will be provided in the following section.

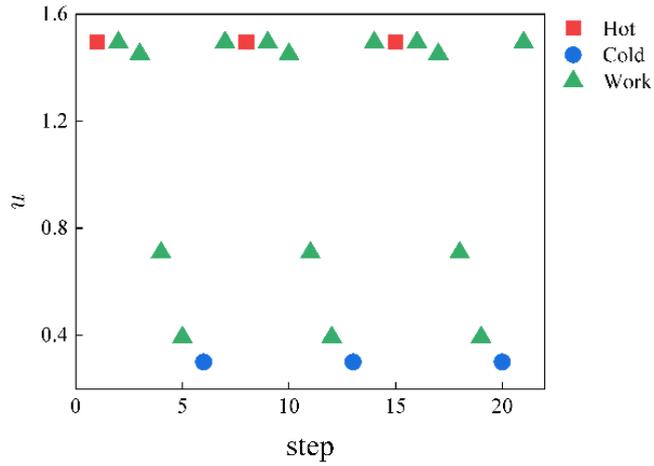

Fig. 5. Optimal cycle produced by the NNs trained by RL algorithm. One step corresponds to $\Delta t$ in Table 3, controlling parameter *u* represents the system state, and the different processes are denoted by different colors. The red, blue, and green markers indicate the "Hot", "Cold", and "Work" processes, respectively.

**3.2 Convergence of the SAC algorithm**

The SAC algorithm incorporates randomness within its operations. For example, the "batch" used for training is randomly sampled from a dynamically changing "replay buffer" [57]. Moreover, the agent's policy is derived from probability sampling of the output probability distribution [78]. Therefore, its stability needs to be scrutinized. Thus, Fig. 6 gives the average power of five consecutive trainings of the coherent three-level QHE. These results demonstrate a similar converged average power of all five trainings, which indicates the reliability of this algorithm.

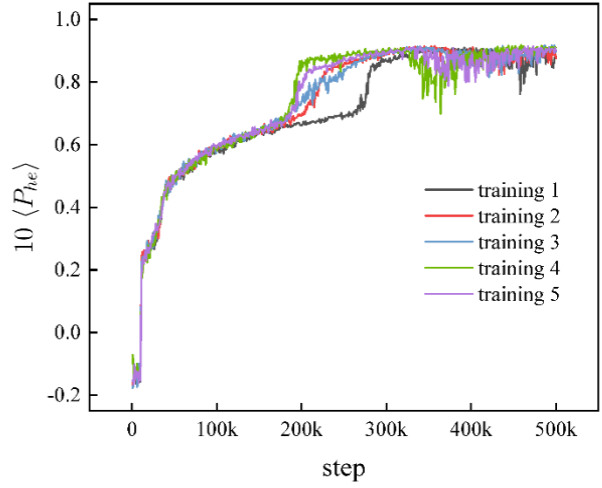

Fig. 6. Averaged power curves of five consecutive trainings for the coherent three-level QHE.

**3.3 Comparative analysis of average power across different cycles**

In the section, the hot process and cold process were unchanged, but the work process was modified based on the cycle given by the RL agent. Fig. 7 (a), (b) and (c) show the patterns of Cycle 1, Cycle 2 and Cycle 3, respectively. Each cycle shares the same hot and cold processes, but the work process varies. The $u$ of work processes in Cycle 1 and 2 changes linearly, with the work process of Cycle 2 enduring longer to achieve sufficient work. Conversely, Cycle 3 adopts the work processes given by the RL agent.

The system is initialized to the Gibbs state,

$$\rho_S(0) = \frac{e^{-\beta_S(0)H_S(0)}}{\mathrm{tr}\left(e^{-\beta_S(0)H_S(0)}\right)} \tag{38}$$

where,

$$\beta_S(0) = 3/\omega_{10}$$
$$H_S(0) = \begin{pmatrix} \omega_0 & 0 & 0 \\ 0 & \omega_1 & 0 \\ 0 & 0 & \omega_2 \end{pmatrix} \tag{39}$$

The final average power over 1k steps, corresponding to the three cycle modes discussed above, is depicted in Fig. 7(d). It can be seen from the figure that the average power of Cycle 1, 2, and 3 are 0.204, 0.591, and 0.837 respectively. This indicates that Cycle 3 given by RL agent, holds a significant advantage of Cycle 1 and 2. Specifically, when compared to Cycle 2, Cycle 3 can improve the performance by approximately 41.6%.

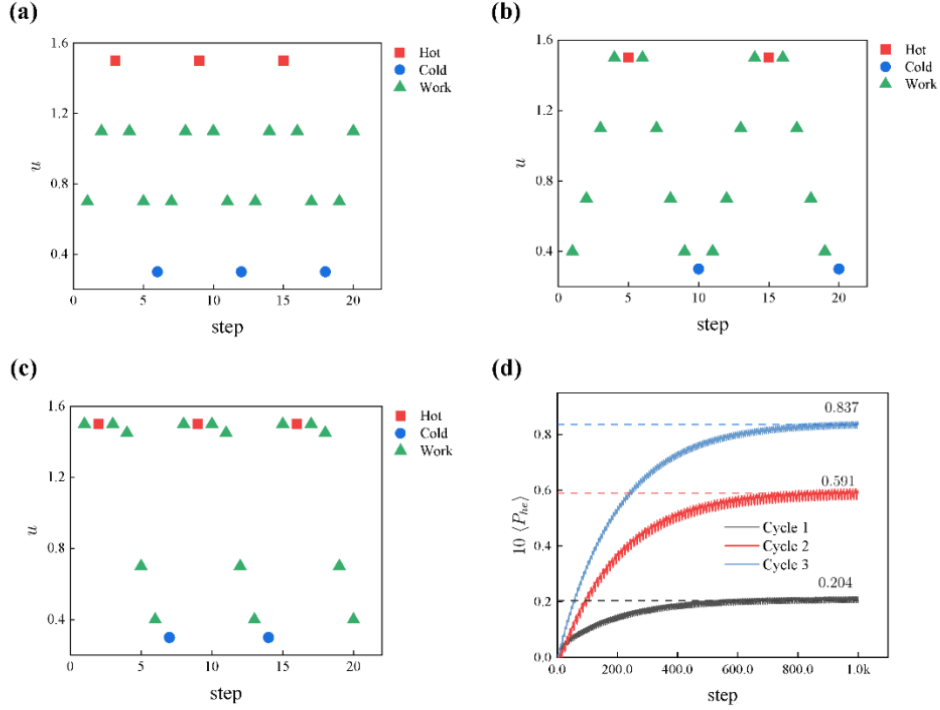

Fig. 7. Different cycle modes and their average power. (a), (b) and (c) represent Cycle 1, Cycle 2 and Cycle 3 as described in the main text, respectively. Each step signifies a time of 0.5. The hot, cold, and work processes are represented by red, blue, and green colors, respectively, while $u$ denotes the system state. (d) depicts the average power under three distinct cycle modes over 1k steps. The black, red, and blue solid lines show the evolution of average power changes under Cycle 1, Cycle 2, and Cycle 3, respectively. The dotted lines indicate the final average power of corresponding cycle.

### 3.4 Efficiency of RL cycle

The quantum system is initialized to the same Gibbs state given in Eq. (38) then evolves for 1k steps according to the RL cycle. After the evolution is over, the cycle efficiency is calculated by the following formula [73]:

$$\eta = \frac{\eta_C}{1 + \frac{\langle \sigma \rangle}{\beta_c \langle P_{he} \rangle}} \tag{40}$$

where, Carnot efficiency $\eta_C = 1 - \beta_h/\beta_c$, average power $\langle P_{he} \rangle$ is given by Eq. (11), and average entropy production,

$$\langle \sigma \rangle = \tilde{\gamma} \int_0^\infty e^{-\tilde{\gamma} t} \sigma(t)\, dt \tag{41}$$

Here, the instantaneous entropy production is given by:

$$\sigma(t) = \sum_{\alpha=c,h} \beta_\alpha J_\alpha(t) \tag{42}$$

According to Eq. (40), the efficiency of the RL cycle can be obtained at approximately 65.4%. This efficiency is greater than the Curzon-Ahlborn efficiency which is the efficiency at maximum power (EMP) derived by Curzon and Ahlborn [81], with $\eta^{EMP} = 1 - \sqrt{\beta_h/\beta_c} = (1 - \sqrt{1/5}) \times 100\% = 55.3\%$. However, recent research [82-85] had shown that the EMP of QHE may exceed $\eta^{EMP}$, which is consistent with our results.

**3.5 Fitting of RL cycle**

The cycle (Fig. 5) obtained by the RL agent in this study can be fitted to an periodic cycle shown in Fig. 8. The duration of working-1, heating, working-2, and cooling process are $\tau_1$, $\tau_2$, $\tau_3$, and $\tau_4$, respectively. As can be observed from the figure that $u$ remains unchanged at 1.495 and 0.300 during the heating and cooling process respectively. Meanwhile, the value of $u$ during the working-2 processes can be fitted by the Boltzmann function [86], as described in Eq. (43) **Error! Reference source not found.**. The specific fitting parameters are provided in Table 4. The coefficient of determination, $R^2$, for the working-2 process is 0.999, demonstrating an approximation of good acceptance. Consequently, we hypothesize that the working-1 process could also be approximated by the Boltzmann function. It should be note that the working-1 process only has one data point within one period, meaning it could be fitted by any function. We attempted to reduce $\Delta t$ to obtain more data points, but the SAC algorithm failed to converge.

$$u(t) = \frac{A_1 - A_2}{1 + e^{(t-t_0)/dt}} + A_2 \qquad (43)$$

$$u(t) = \frac{A_1 - A_2}{21^{(t-t_0)/dt}} + A_2$$

Table 4. Parameter values of the fitted Boltzmann function for different processes. $t$ denotes the duration of the process with a unity of $\Delta t$, while $A_1$, $A_2$, $t_0$, and $dt$ are the fitted Boltzmann function parameters, $R^2$ is the coefficient of determination. The last row displays the parameters of the general working process, where the values outside and inside the brackets refer to the working-1 and working-2 process, respectively.

| Process | $t$ | $A_1$ | $A_2$ | $t_0$ | $dt$ | $R^2$ |
|---|---|---|---|---|---|---|
| working-1 | [6, 7.5] | 0.300 | 1.495 | 6.75 | 0.05 | |
| working-2 | [8.5, 12] | 1.497 | 0.300 | 10.25 | 0.25 | 0.990 |
| working | $[t_{\min}, t_{\max}]$ | 0.3(1.5) | 1.5(0.3) | $\dfrac{t_{\min}+t_{\max}}{2}$ | 0.05(0.25) | |

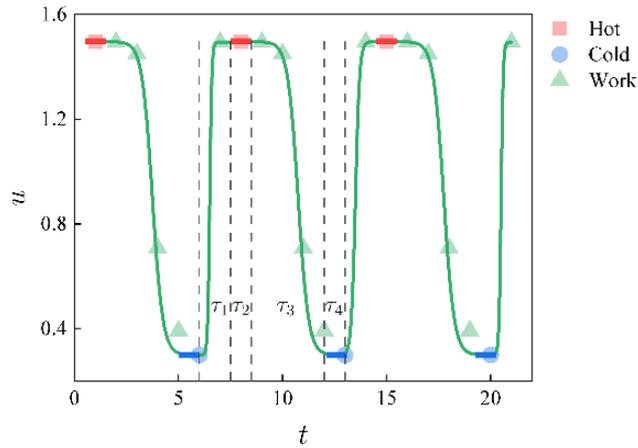

Fig. 8. Fitted cycle based on the RL cycle depicted in Fig. 5. The red, blue, and green solid lines represent the heating, cooling, and working processes, respectively. The translucent markers indicate the RL cycle. A single cycle of the fitted Otto cycle is marked by dashed lines, with $\tau_1$, $\tau_2$, $\tau_3$, and $\tau_4$ represent the duration of working-1, heating, working-2, and cooling process, respectively.

This fitted cycle can be regarded as an Otto-like cycle. The quantum Otto cycle generally includes four processes: adiabatic compression, isochoric heating, adiabatic expansion, and isochoric cooling [87]. The working-1 and working-2 process respectively increases and decreases the energy gap, which are similar to the compression and expansion process within the quantum Otto cycle. Therefore, this fitted cycle is analog to the Otto cycle for they both maximize the power, demonstrating the effectiveness of our method.

**3.6 Discussion with the finite-time Otto cycle**

Applying the RL algorithm also provides a new perspective for investigating the finite-time Otto cycle with the advantage of alleviating tedious analysis. The finite-time Otto cycle [29, 37, 38, 44-54, 88] is proposed to deal with the practical application defects of the ideal Otto cycle. The ideal Otto cycle generally needs to meet two assumptions. Firstly, the compression and expansion process should be quasi-static to prevent heat leakage. Secondly, the system should be in the Gibbs state after the isochoric processes. However, in actual processes, the quasi-static processes and the slow isochoric processes required by the Gibbs state could lead to a decreased power. To this end, some finite-time Otto cycles, such as utilizing "adiabatic shortcut" [48-54] to speed up the adiabatic processes, are designed to improve output power. However, most of these shortcuts rely on experience or require complex theoretical derivations which hinder the corresponding research.

It should be noted that due to the complexity of theoretical derivation and calculation, research on the finite-time Otto cycle of the three-level QHE is currently challenging and limited [4, 46, 76, 82, 89]. Thus, this study only compares the power of the steady-state and does not delve into the power of the Otto cycle.

## 4 Conclusions

This research employed the RL by the SAC algorithm to optimize the long-term performance of the coherent three-level quantum heat engine (QHE), specifically aiming to maximize the average output power. Remarkably, the RL agent gave a cycle with an average output power of approximately 1.28 greater than the steady limit. Furthermore, the convergence of the SAC algorithm was verified through five consecutive trainings and the efficiency of this cycle was found to be larger than the Curzon-Ahlborn efficiency. Finally, the results also showed that the optimal cycle could be fitted as an Otto-like cycle by adopting the Boltzmann function during the compression and expansion processes. This demonstrates the feasibility of utilizing reinforcement learning within the power optimization of QHE.

## Acknowledgement

This work was supported by the Taishan Scholar Project (Grand No. tsqn202103142), Natural Science Foundation of Shandong Province (No. ZR2021QE033).

## Competing Interests

The authors declare that they have no known competing financial interests or personal relationships that could have appeared to influence the work reported in this paper.

## Dada Availability

The data that support the findings of this study are available from the corresponding author upon reasonable request.

## Author contributions

Gao-xiang Deng, Wei Shao and Zheng Cui designed this research and its corresponding model and method. Gao-xiang Deng wrote the code, carried out the training and processed the data. Gao-xiang Deng, Haoqiang Ai, Bingcheng Wang, Wei Shao and Yu Liu wrote the manuscript.